\documentclass[final,5p,times,twocolumn]{elsarticle}

\usepackage{graphicx}

\usepackage{amssymb}

\usepackage{amsmath}

\usepackage{lineno}

\journal{Nuclear Instruments and Methods in Physics Research A}

\begin{document}

\begin{frontmatter}



\title{Two-dimensional visualization of cluster beams by microchannel plates} 


\author[adr1]{A. Khoukaz\corref{cor}}
\ead{khoukaz@uni-muenster.de}
\cortext[cor]{Corresponding author.}

\author[adr1]{D. Bonaventura}
\author[adr1]{S. Grieser}
\author[adr1]{A.-K. Hergem\"oller}
\author[adr1]{E. K\"ohler}
\author[adr1]{A. T\"aschner}

\address[adr1]{Institut f\"ur Kernphysik, Universit\"at M\"unster, D-48149 M\"unster, Germany}

\begin{abstract}
An advanced technique for a two-dimensional real time visualization
of cluster beams in vacuum as well as of the overlap volume of 
cluster beams with particle accelerator beams is presented. 
The detection system consists of 
an array of microchannel plates (MCP) in combination with a phosphor screen which is read
out by a CCD camera. 
This setup together with the ionization of a cluster beam by an electron or ion beam allows for spatial resolved investigations of the cluster beam position, size, and intensity. Moreover, since electrically uncharged clusters remain undetected, the operation in an internal beam experiment opens the way to monitor the overlap region and thus the position and size of an accelerator beam crossing an originally electrically neutral cluster jet.
The observed intensity distribution of the recorded image is directly 
proportional to the convolution of the spatial ion beam and cluster beam intensities and
is by this a direct measure of the two-dimensional luminosity distribution. This
information can directly be used for the reconstruction of vertex positions as well
as for an input for numerical simulations of the reaction zone. The spatial resolution 
of the images are dominated by the granularity of the complete MCP device and was
found to be in the order of $\sigma\,\approx\,100\,\mu\mathrm{m}$. 
\end{abstract}


\end{frontmatter}


\section{Introduction}
Cluster beams can be produced and prepared from originally solid, liquid or gaseous 
materials and are widely used in modern physics experiments. Prominent
examples for this are cluster jet beams as internal targets for storage ring experiments
\cite{Eks95,Dom97,Allspach1998,Tas11} or interaction studies of high-intense laser beams 
with streams of clusters \cite{Li06}.
Of special interest are cluster beams produced via expansion of cooled gases or liquids 
in Laval nozzles by either condensation of the gas or by breaking
up the liquid into a spray of droplets. Depending on the production parameters 
these clusters typically consist of $10^3-10^6$ molecules~\cite{Knuth1995,General2008}. 
Target streams for experiments can be provided with freely adjustable target thickness over 
several orders of magnitude up to about $\rho_{\mathrm{target}}=10^{15}\,\mathrm{atoms/cm}^2$
in a distance of more than $2\,\mathrm{m}$ behind the nozzle \cite{Tas11}. 
Due to the high mass of the individual clusters compared
to the residual gas background they can travel over several meters 
through an ultra high vacuum chamber with a constant angular divergence defined by the
orifices used for the cluster beam preparation. This fact is of high 
interest if the experimental setup,  e.g. of a cluster target at an accelerator facility, 
requires large distances between the cluster generator and both the scattering chamber and 
the cluster beam dump. Especially in this case with large distances a careful cluster target 
beam alignment is mandatory and suitable devices for this are of interest. 

\begin{figure}
\includegraphics[width=\columnwidth]{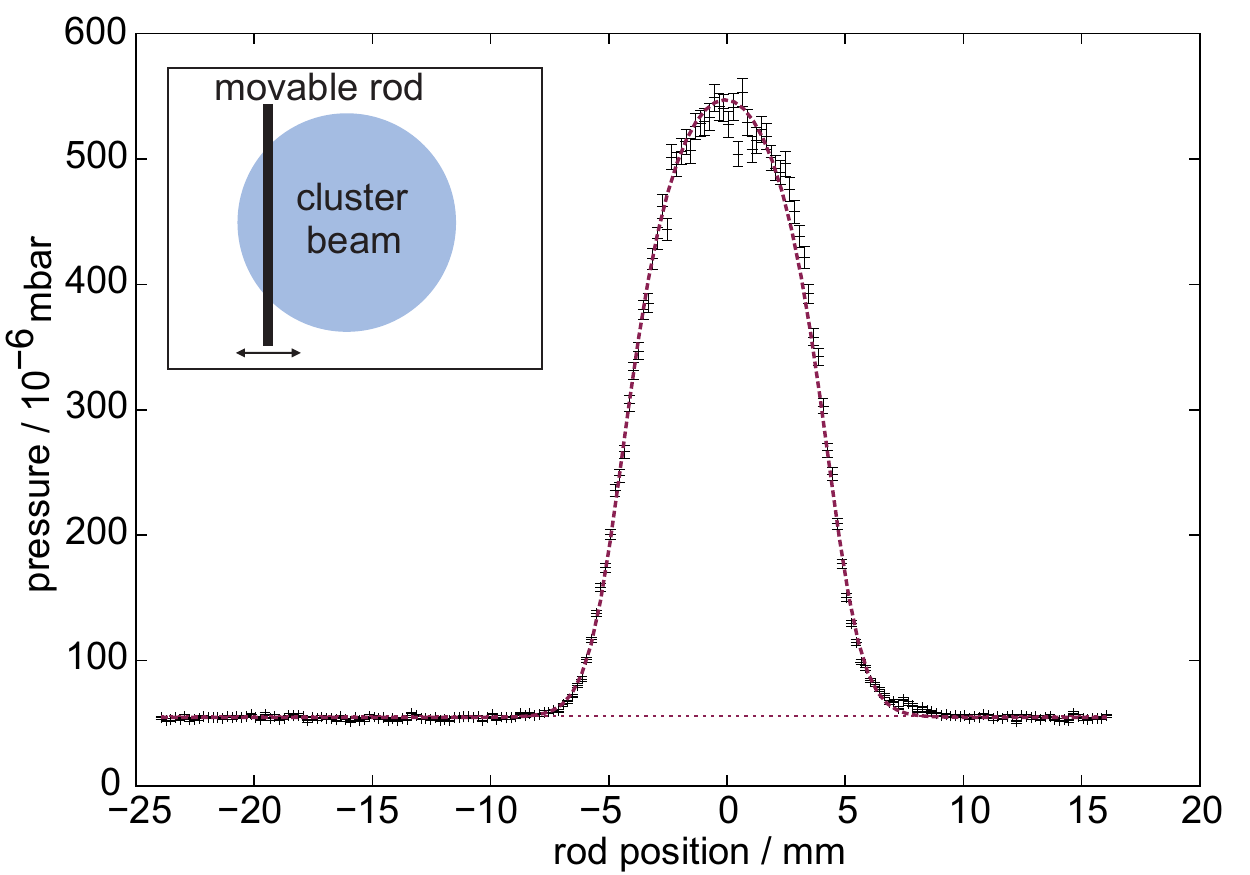}
\caption{Determination of the size and intensity of a cluster beam by the scanning rod method.}
\label{fig:scanrod}
\end{figure}

A further important quantity in scattering experiments using cluster beams as 
targets, e.g. for accelerated ion beams, is the spatial distribution of the interaction region
as well as its variation with time. 
Here a precise knowledge of this information might be needed, e.g. for the
reconstruction of the tracks of the ejectiles produced in the interaction volume or 
for realistic Monte-Carlo computer simulations for the reactions of interest.
One possibility to gain such information about the vertex volume might be the investigation of 
fluorescence light produced by excitation of atoms/molecules of the target beams 
by the passing ions.
However, this method commonly requires an optical access to the interaction point 
which might be difficult or even not be possible for modern compact $4\pi$-detectors 
such as the planned $\bar{\mathrm{P}}$ANDA detector at FAIR in Darmstadt \cite{PANDA2005}.
Alternatively the required vertex information might be reconstructed by separate 
investigations on the accelerator beam profile, e.g. using a fluorescence light
measurements \cite{Satou2006,Tsang2008} of (residual) gas close to the interaction region, 
in combination with a measurement on the cluster target beam profile. 
In order to quantify the geometrical size and/or the intensity of a cluster beam at the 
interaction point of an experiment, different approaches can be followed. One commonly applied 
technique for cluster beams from originally gaseous materials is the use of thin scanning rods 
in one of the differentially pumped vacuum stages. 
The principle of this method is shown in Fig.~\ref{fig:scanrod}.

A movable rod scans the cluster beam in the vacuum chamber and as soon as the rod interacts with the 
cluster stream the clusters of the overlap region are stopped 
and converted into an additional gas background. The resulting gas load to the vacuum chamber can
easily be measured by vacuum gauges and is directly proportional to the cluster beam density.
Furthermore, with the known rod dimensions and the recorded pressure rise as a function of the rod 
position, information about the cluster beam diameter and the local target thickness can be 
extracted \cite{Tas11}.
With this method only one-dimensional 
scans are possible which result in data averaged over the rod axis. In addition, depending on the
used vacuum gauge and its possible read out speed, one single cluster profile measurement might 
take, e.g. one minute, if a sufficient spatial resolution is required. 
To overcome the previously described limitations, a method based on a microchannel plate device 
(MCP) has been implemented. It allows for a precise cluster target beam position measurement 
and adjustment, to measure in real time two-dimensional cluster beam intensity distributions as 
well as to monitor the spatial distribution of the vertex point if the cluster target is used in
combination with an accelerator beam. 

The principle of operation and the properties of this device are demonstrated here by data obtained 
with hydrogen cluster beams. While electrically neutral clusters shot directly onto a 
MCP device remain undetected, ionized clusters, e.g. produced by electron impact, can be registered 
by this system. 
Therefore, the complete cluster beam cross section can be investigated. 
If the MCP signal is read out by a phosphor screen and a CCD camera, the time needed to
measure one cluster beam image is typically only limited by the required exposure time
of the CCD camera in order to collect a sufficient amount of photons. However, this time
might be reduced,  e.g. to a few seconds, if the electron current for cluster ionization is
adjusted correspondingly. The presented system
is routinely in operation as diagnostic system at a hydrogen cluster beam installation
at the University of M\"unster. 
Furthermore, with this device it was possible to visualize and monitor the interaction 
region of a hydrogen cluster target beam and a proton beam in a storage ring experiment. 
In this case the ionization of the originally electrically neutral cluster beam resulted from 
the energy loss processes of the ion beam in the cluster beam. 

\section{Experimental setup}

\begin{figure}[h]
\begin{center}
\includegraphics[width=0.8\columnwidth]{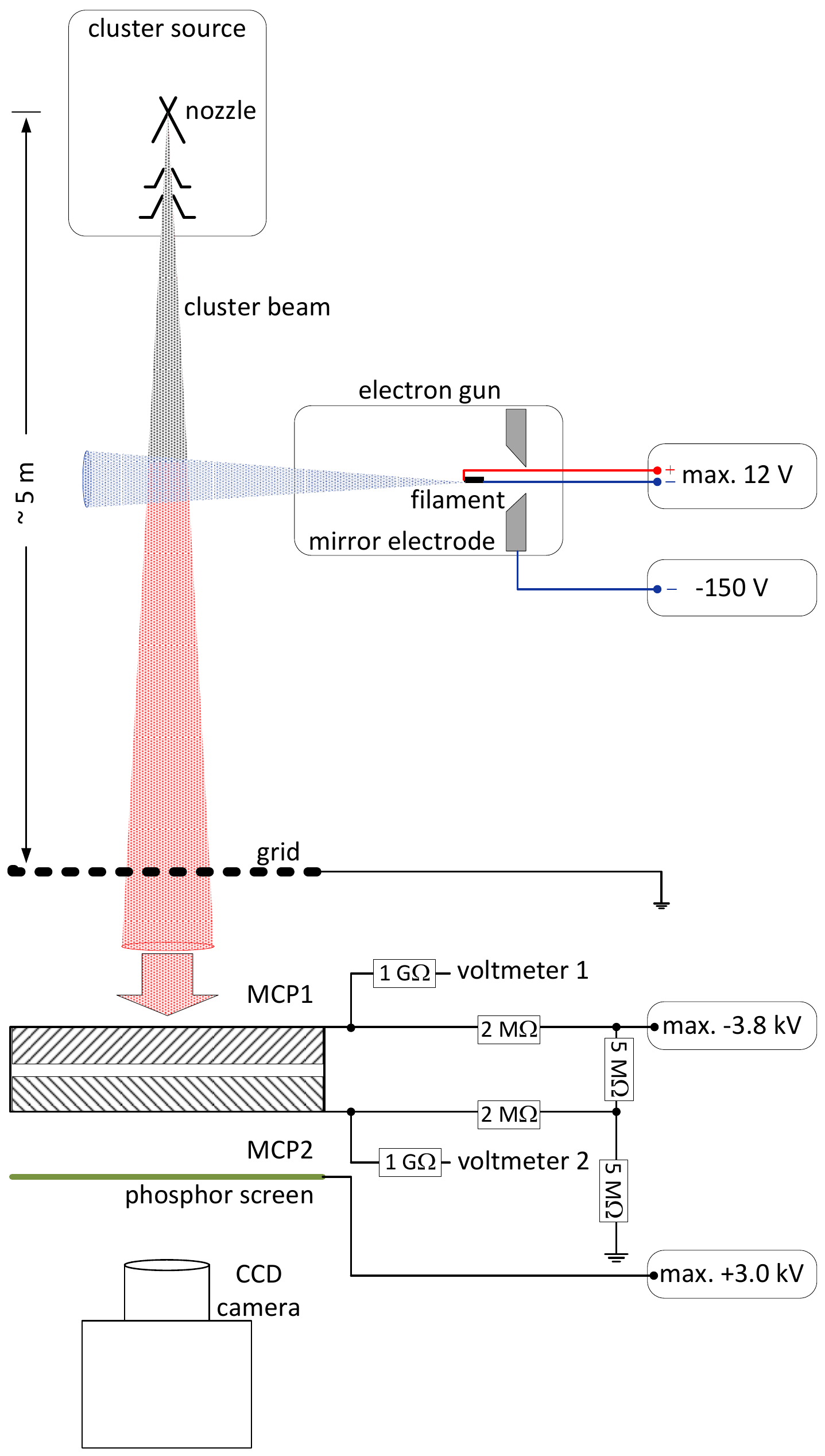}
\end{center}
\caption{Schematic view of the MCP based setup for the detection of ionized cluster beams.}
\label{fig:mcp1}
\end{figure}

A hydrogen cluster beam is produced via adiabatic expansion of pre-cooled and compressed hydrogen 
in a Laval nozzle. Details about the cluster jet generator, the experimental setup as well as 
the operational
parameters are presented in Ref.\,\cite{Tas11}. Shortly behind the cluster generator the cluster 
beam is ionized by an electron gun which can be operated in a continuous mode or, if a timing 
information is needed, in a pulsed mode. At the crossing point of cluster and electron beams, i.e. 
$76\,\mathrm{cm}$  
behind the nozzle, the diameter of the electron beam amounts to approximately $7\,\mathrm{mm}$  
and is larger than the cluster beam, i.e. 4\,mm, at that position. By this a complete coverage of the 
hydrogen beam is guaranteed. The current of the $150\,\mathrm{eV}$ electron beam is adjusted to the 
experimental needs, e.g. to the required intensity of ionized clusters, and is typically in 
the order of a few microampere. At the interaction point 
clusters with both negative and positive charges are produced from which in the following only the 
positively charged ones are further considered. 
After a drift path of approximately $4.2\,\mathrm{m}$, measured from the position 
of the electron beam, 
the ionized beam is shot onto the MCP based detection system.  
A schematic view of this device as well as the electrical circuit is shown in Fig.~\ref{fig:mcp1}.

The cluster beam, indicated by a vertical cone, passes an electrically grounded grid with a spacing of 
$2.7\,\mathrm{mm}$ and hits a microchannel plate in Chevron assembly. While depending on the operational 
parameters of the cluster jet generator the individual clusters have originally velocities in the order of
$200-1000\,\mathrm{m/s}$ \cite{Tas11,Koehler2010,TaeschnerDr} before entering the detector, 
the positively electrically charged 
clusters are 
accelerated after passing the grid due to the potential of up to $-3.8\,\mathrm{kV}$ at the entrance 
surface of the MCP.
By this the kinetic energy of the clusters is high enough to produce primary electrons which can be 
multiplied within the MCP device. The cluster beam itself is stopped at the surface of the MCP, converted
into residual gas and is pumped away by a turbomolecular pump. Directly behind the MCP a phosphor screen at a potential of $+3.0\,\mathrm{kV}$ is placed which is hit by the electrons from the MCP. The resulting image
can be observed by a CCD camera behind an UHV window outside of the vacuum.

\begin{figure}[h]
\includegraphics[width=\columnwidth]{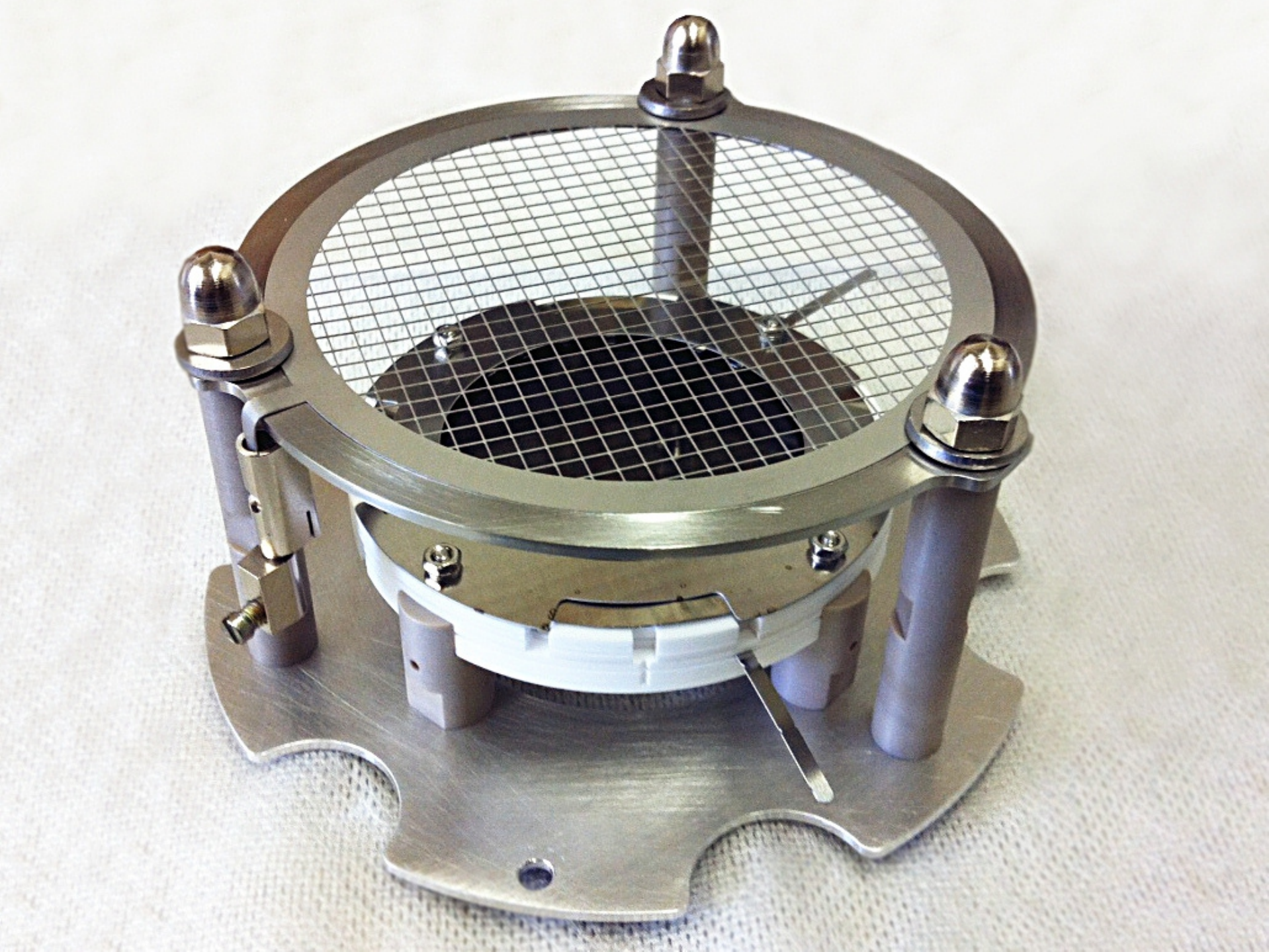}
\caption{Photography of the MCP device with the entrance grid. The ionized cluster beam enters 
from the top and hits the MCP array.}
\label{fig:mcpphoto}
\end{figure}

In Fig.~\ref{fig:mcpphoto} a photography of the MCP device with the entrance grid is shown. 
The MCP (dark area below the grid) with a gain of $>4\times 10^6$ has an active diameter 
of $40\,\mathrm{mm}$ and a pore diameter of $(12.0\pm0.5)\,\mu\mathrm{m}$. The adapted phosphor screen is of the type P43 
which produces photons of green color. The entrance grid was produced by a commercial etching procedure 
with a bar width of $0.2\,\mathrm{mm}$. 

\begin{figure}
\includegraphics[width=\columnwidth]{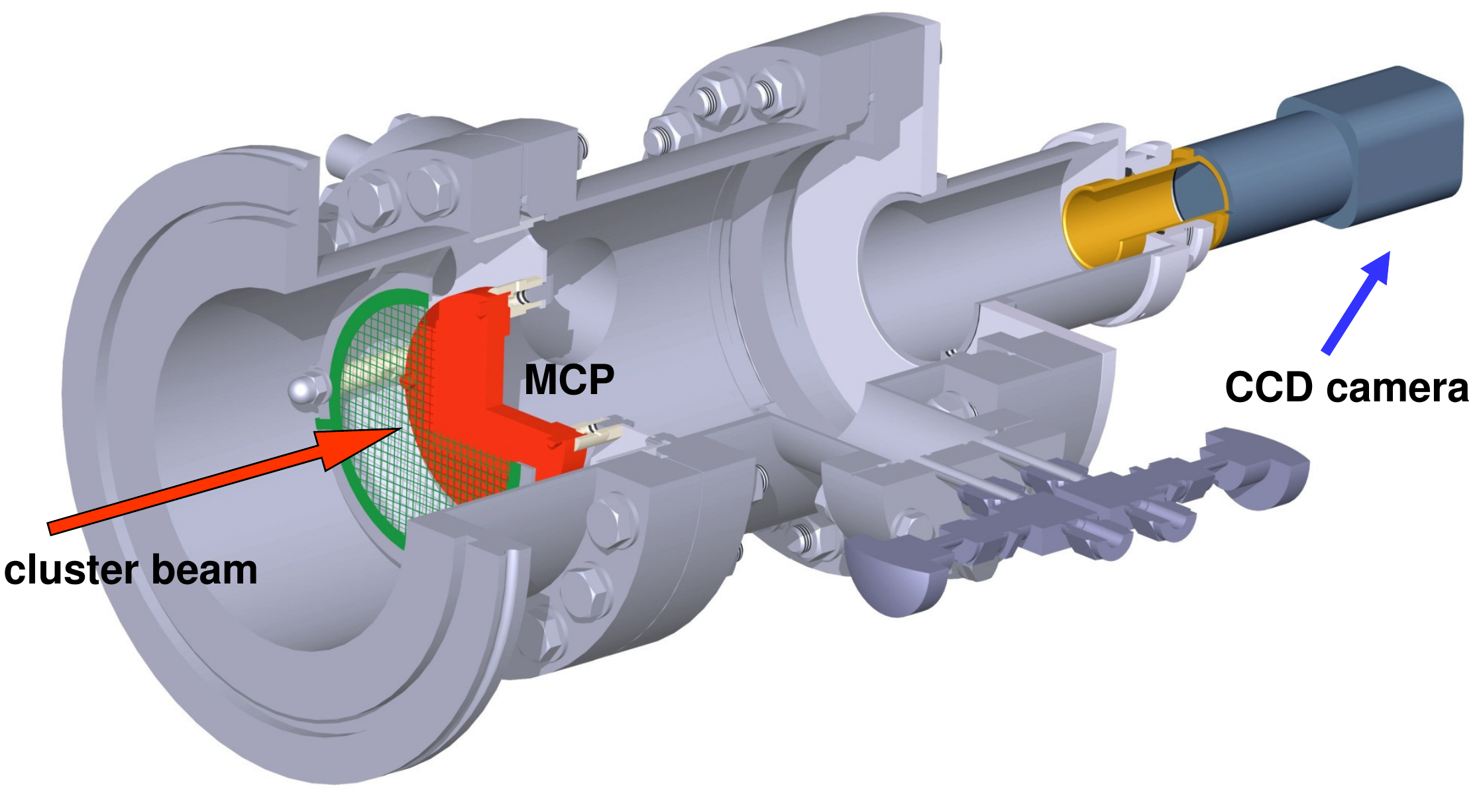}
\caption{Schematic view of the complete MCP detection system with the CCD camera.}
\label{fig:mcpfull}
\end{figure}

In Fig.~\ref{fig:mcpfull} a schematic view of the complete setup is displayed, 
showing the final vacuum stage with the MCP detection system as well as the CCD camera.

\section{Visualization of cluster beams}
\begin{figure}
\begin{center}
\includegraphics[width=0.75\columnwidth]{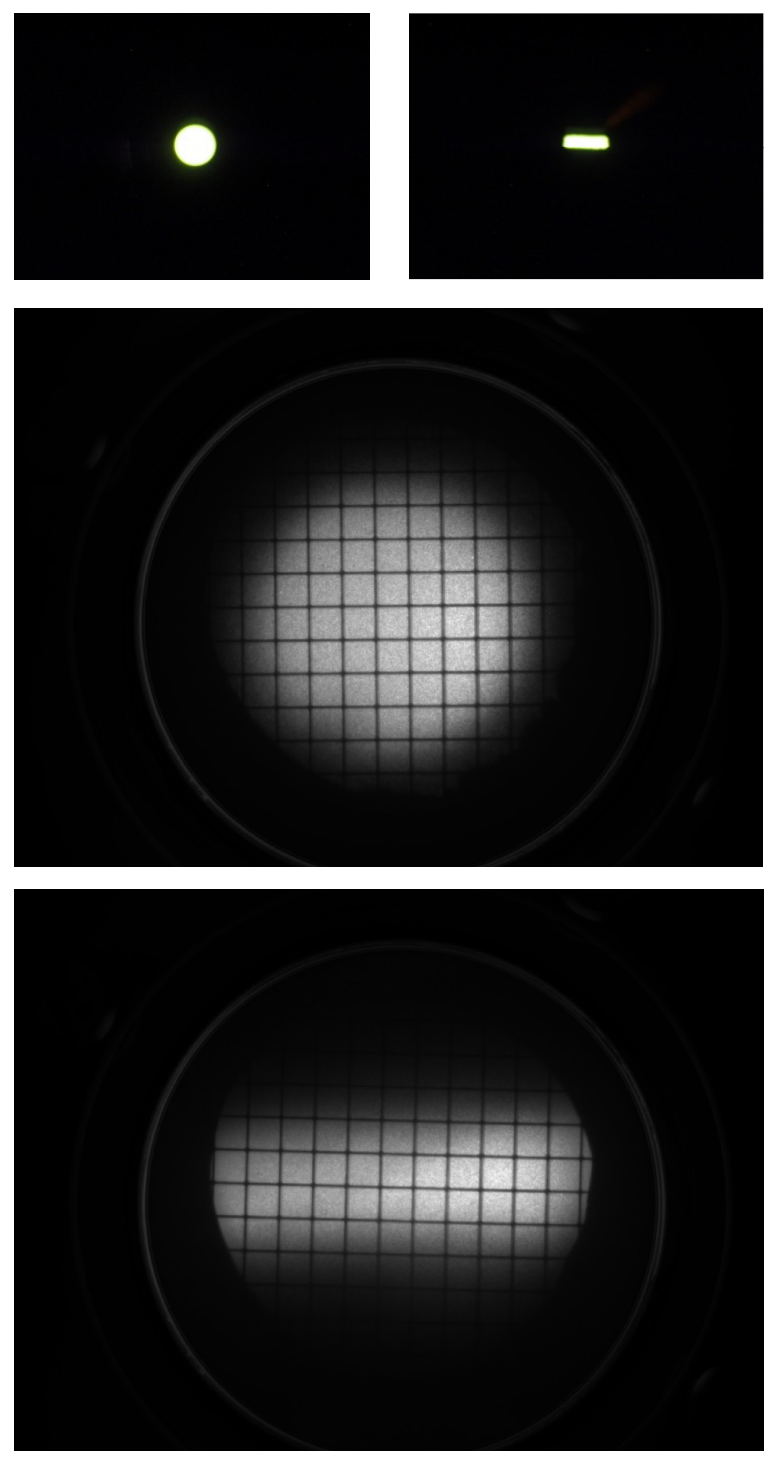}
\end{center}
\caption{Microscopic view of two collimators (top) with round ($\varnothing \approx\mathrm{0.5\,mm}$) and 
slit ($\mathrm{0.77\times0.19\,mm^{2}}$) shaped orifices and resulting MCP images (bottom) of the 
cluster jet beam at a distance of approximately $5\mathrm{\,m}$ behind the nozzle.}
\label{fig:twoclusterbeams}
\end{figure}

To investigate the possibility to measure two-dimensional cluster beam density profiles,
cluster beams with different geometries have been prepared by using special shaped collimators,
placed $\mathrm{121\,mm}$ behind the nozzle \cite{Her2013}. 
In Fig.~\ref{fig:twoclusterbeams} (top) microscopic views of two different collimators are
shown which have been fabricated by laser cutting of stainless steel cones. 
The constant angular divergences of the cluster beams passing these orifices
are determined by the size of the 
collimator openings as well as the distance between the nozzle and the collimator.
This in turn determines the sizes of the cluster beams at the following vacuum chambers
as well as at the position of the MCP assembly.
The resulting images recorded by the MCP system are presented by the lower pictures of 
Fig.~\ref{fig:twoclusterbeams} \cite{GSIa}. The clearly visible grid spacing allows for a direct measurement
of the cluster beam size at that distance from the nozzle, i.e. after approximately $5\,\mathrm{m}$.
While the round shaped cluster beam, prepared by a collimator with a diameter of $500\,\mathrm{\mu m}$, 
is fully visible by the MCP system, the rectangular shaped beam is not fully detected in one direction
due to the use of a collimator with a width of $190 \times 770\,\mathrm{\mu m^2}$. By this a part 
of the large cluster beam is cut away by an orifice inside of the beam dump stage which is located 
directly 
in front of the MCP system. The shape of this internal orifice is clearly visible. Thus this
kind of detector allows also for an online monitoring of the target beam adjustment and interferences, 
e.g. with vacuum components.
The observed boundary smearing 
of the cluster beam is in full agreement with earlier findings using the scanning method with 
moveable rods (see Fig.~\ref{fig:scanrod}).

\begin{figure}
\includegraphics[width=\columnwidth]{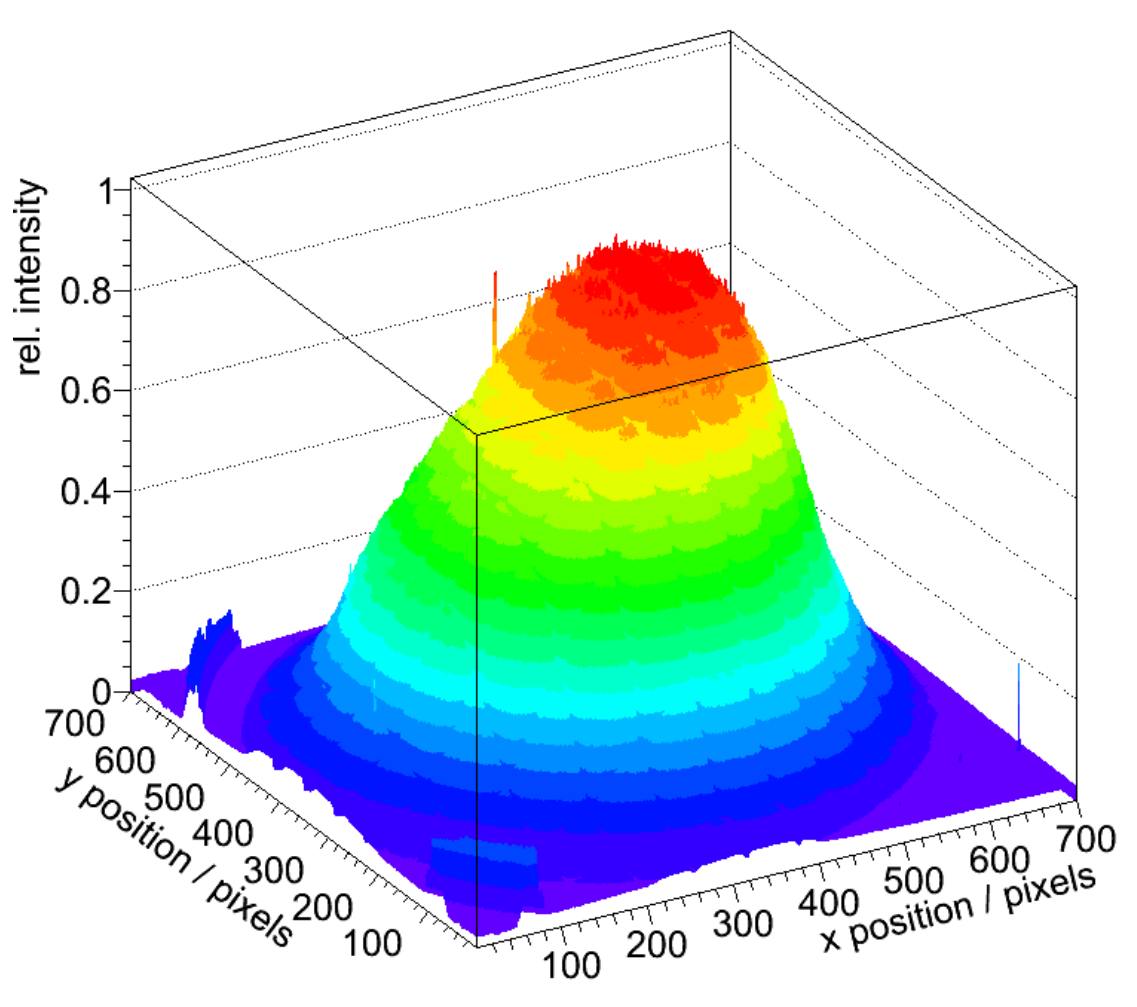}
\caption{Two-dimensional intensity distribution of a cluster beam obtained by the MCP device.}
\label{fig:mcp3d}
\end{figure}

For a more quantitative analysis of the recorded figures the two-dimensional intensity distributions
have been analyzed. Fig.~\ref{fig:mcp3d} shows exemplary the result for the circular cluster beam
shown in Fig.~\ref{fig:twoclusterbeams}. Here the $x$- and $y$-axes correspond to the directions
perpendicular to the spread direction of the cluster beam. By using a homogeneously distributed 
electron beam for the cluster ionization, the intensity distribution observed with the MCP device
is a direct measure for the local cluster beam intensity. A closer investigation of the intensity 
distribution exposes a slight asymmetry which might be caused by the not yet perfectly adjusted relative 
position of the two collimators in the cluster source. Therefore, the MCP device presented here is
highly suited for online collimator adjustment purposes during the preparation of cluster beams, 
e.g. for hadron physics experiments.  

\begin{figure}
\includegraphics[width=\columnwidth]{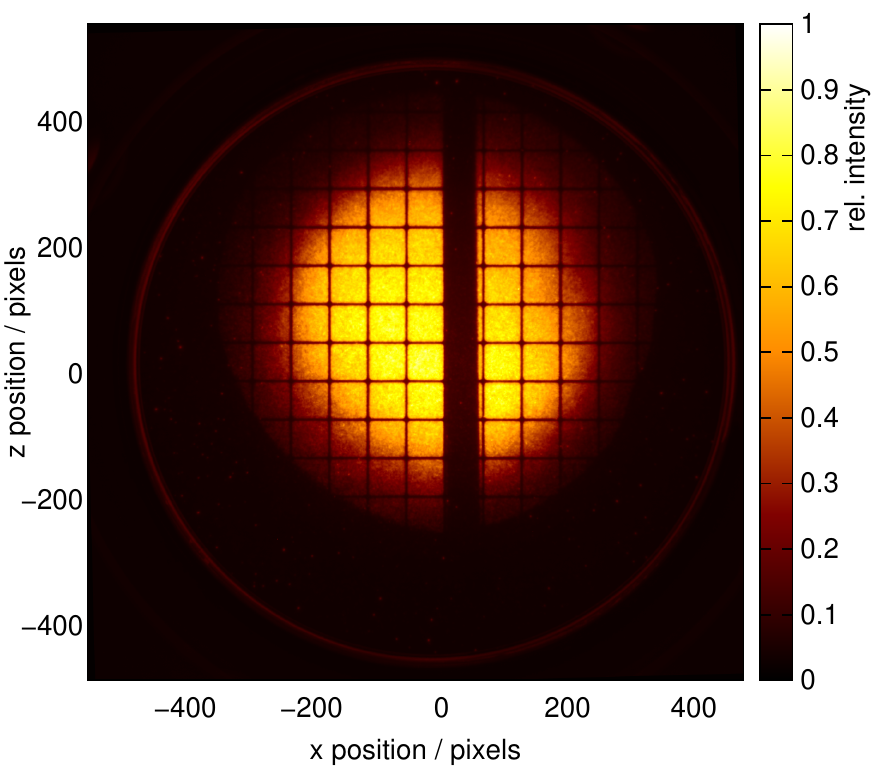}
\includegraphics[width=\columnwidth]{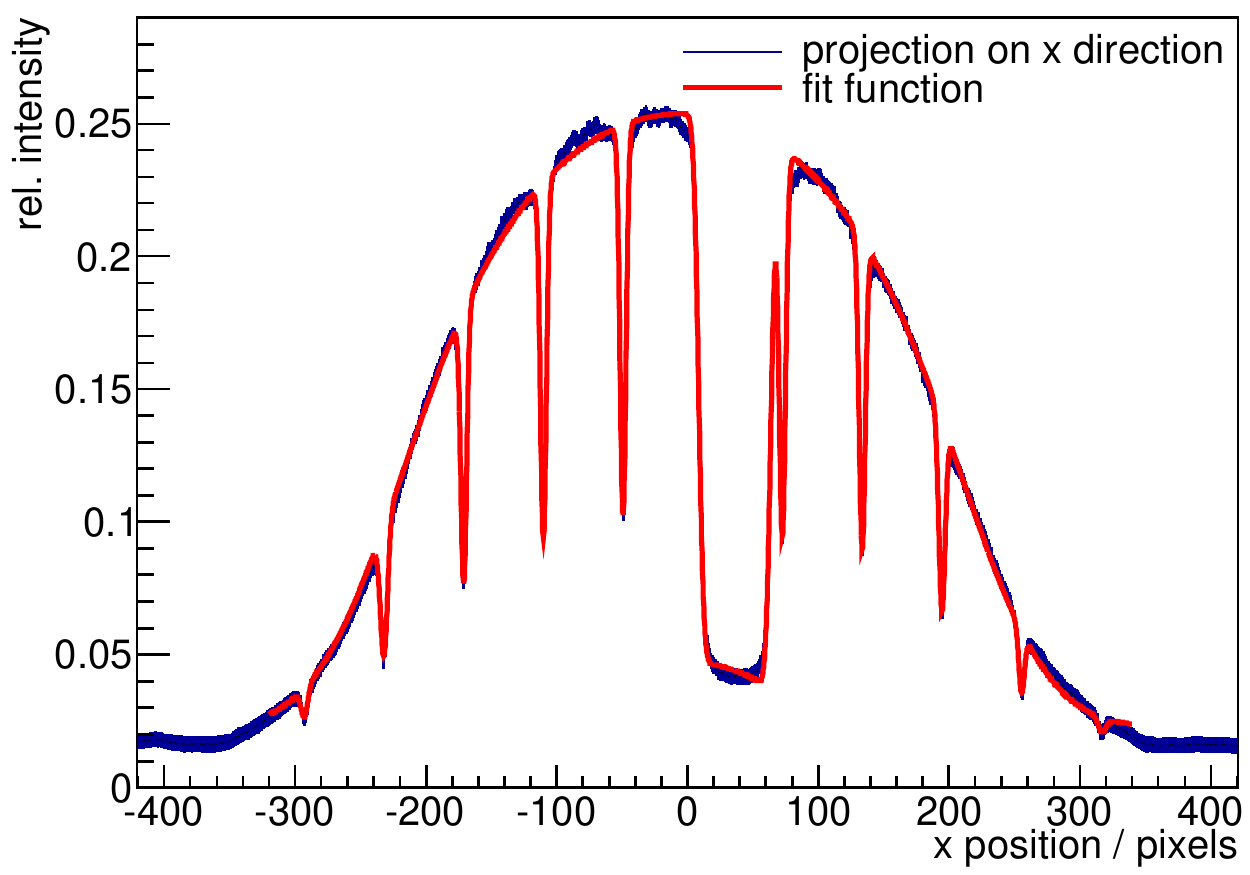}
\caption{MCP image (top) and projection (bottom) of the intensity distribution obtained with one scanning rod 
crossing the cluster beam flight path. }
\label{fig:stab}
\end{figure}

To gain further information about the quality of the spatial resolution of the MCP system,
one of the scanning rods of the monitor system with a diameter of $d_{\mathrm{rod}} = 1 \mathrm{mm}$ located 
at the scattering chamber, i.e. $l_{\mathrm{scat}} = 2.1\, \mathrm{m}$ behind the nozzle, was placed at a fixed position
within the passing cluster beam. Fig.~\ref{fig:stab} (top) shows the extracted MCP image
obtained at a distance of  $l_{\mathrm{MCP}} = 
5.0\, \mathrm{m}$ behind the nozzle. In addition to the already described
cluster beam image a sharp-edged shadow of the rod is clearly visible. To extract information about the 
spatial resolution of the MCP device, a projection of this two-dimensional intensity image was
determined and is displayed in Fig.~\ref{fig:stab} (bottom).
The recorded data points are given by the blue line and the shape of the envelope 
is in good agreement with the expected shape of a profile of a circular cluster beam 
known from the scanning rod method (see Fig.~\ref{fig:scanrod}). The narrow
intensity drops are a result of the regular grid structure and by this an absolute spatial
calibration of the MCP monitor was possible. 
\begin{figure}
\includegraphics[width=\columnwidth]{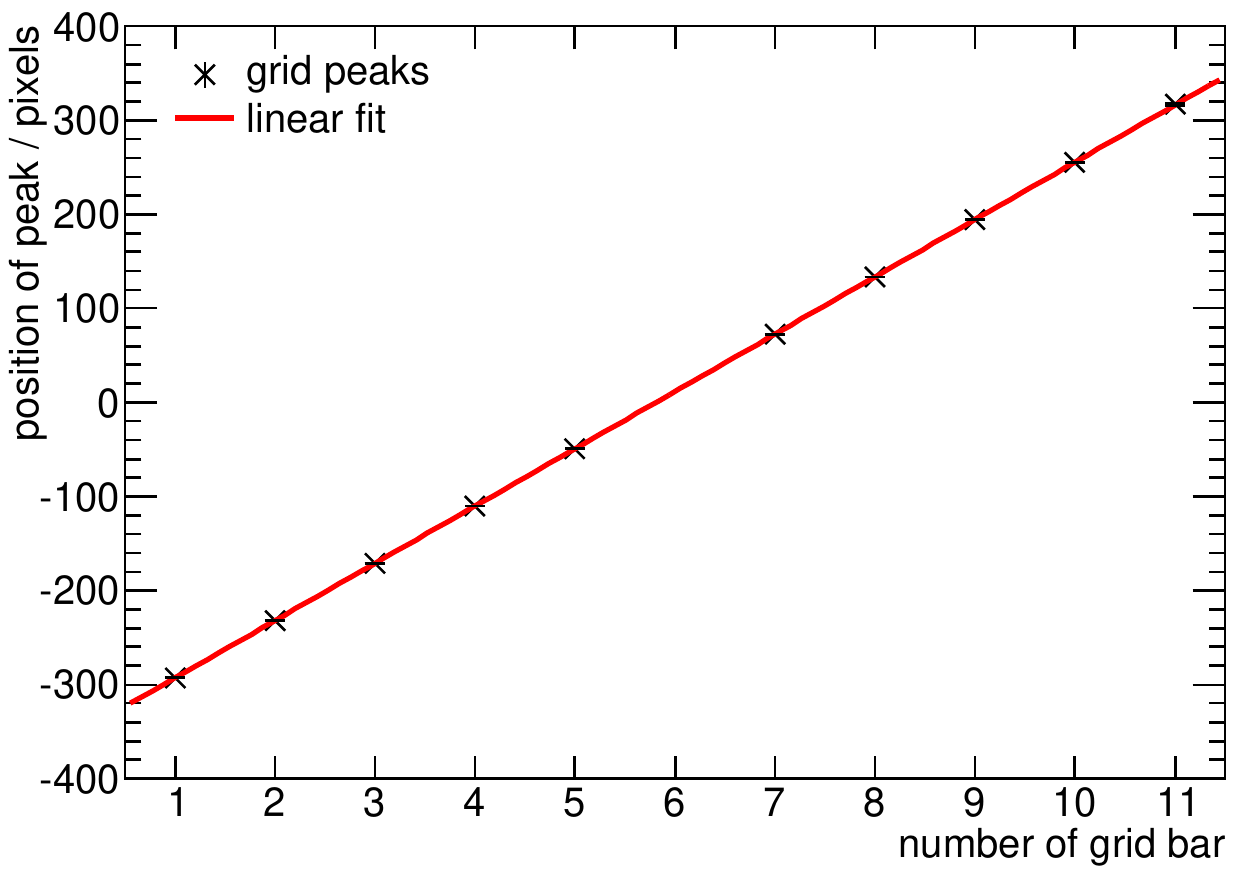}
\caption{Measured position in units of the CCD camera pixels of the intensity drops  
caused by a shadowing effect of the grid bars. }
\label{fig:skala}
\end{figure}
In Fig.~\ref{fig:skala} the measured position in units of the CCD camera pixels of 
the intensity drops is shown. As expected the data can be described well by
a first order polynomial function and expose by this no signal for a non-linearity.    
With the pixel information of the CCD camera and 
the known grid bar distance of $d_{\mathrm{grid}} =2.7\,\mathrm{mm}$ (center-to-center) 
the calibration factor was determined to
be $c_{\mathrm{MCP}} =(44.32\pm0.01)\,\mathrm{pixel/mm}$. 
Moreover, since the shape of the narrow 
intensity drops is given by a convolution of the known grid bar width of $d_{\mathrm{bar}} =200\,\mu\mathrm{m}$ 
and the intrinsic resolution of the MCP detector itself, the latter information can be 
extracted from a fit to the displayed data. An excellent description of the intensity drops
was achieved by a pure Gaussian distribution with a width of $\sigma_{\mathrm{MCP}} =
(103\pm 1)\,\mu\mathrm{m}$ (RMS),
representing an upper limit for the intrinsic resolution of the MCP detector.
Furthermore, the broad minimum visible in Fig.~\ref{fig:stab} (bottom) is a result of the scanning rod
placed inside of the cluster beam. While the total width allows for a reconstruction of the 
rod size, an even more interesting information is given by the shape of the steep intensity 
drops from the rod itself. 
Again a good description of the data was possible by a convolution of a 
rectangular rod geometry and a Gaussian detection resolution $\sigma_{\mathrm{res}}$. Here it must be noted 
that the latter resolution combines contributions from the already determined intrinsic resolution of 
the MCP detector, i.e. $\sigma_{\mathrm{MCP}}$, and additional effects such as possible stray fields influencing the 
trajectories of the ionized cluster beam or minor divergences of the cluster beam itself. 
The best description of the data is shown by the red line and considers a detection resolution 
of $\sigma_{\mathrm{res}}=(135\pm 3)\,\mu\mathrm{m}$ (RMS) in addition to the known
numbers for $d_{\mathrm{grid}}$, $d_{\mathrm{bar}}$, $\sigma_{\mathrm{MCP}}$. Thus, the resolution for the detection of structures
in the cluster beam or of a possible displacement at the position of the scattering chamber, i.e. 
$2.1\,\mathrm{m}$ behind the nozzle, can be estimated to be better than 
\begin{equation}
\sigma_{\mathrm{scat}} \le   \sigma_{\mathrm{res}} \cdot \frac{l_{\mathrm{scat}}}{l_{\mathrm{MCP}}} =  (57 \pm 1)\,\mu\mathrm{m}.
\end{equation}
Therefore, the MCP monitor device presented here is highly suited for the observation and the quantitative 
investigations of cluster beams used as target beams, e.g. in hadron physics experiments at storage
rings or at high power laser facilities.

\section{Visualization of the beam-target vertex region in accelerator experiments}

\begin{figure}[h]
\begin{center}
\includegraphics[width=0.73\columnwidth]{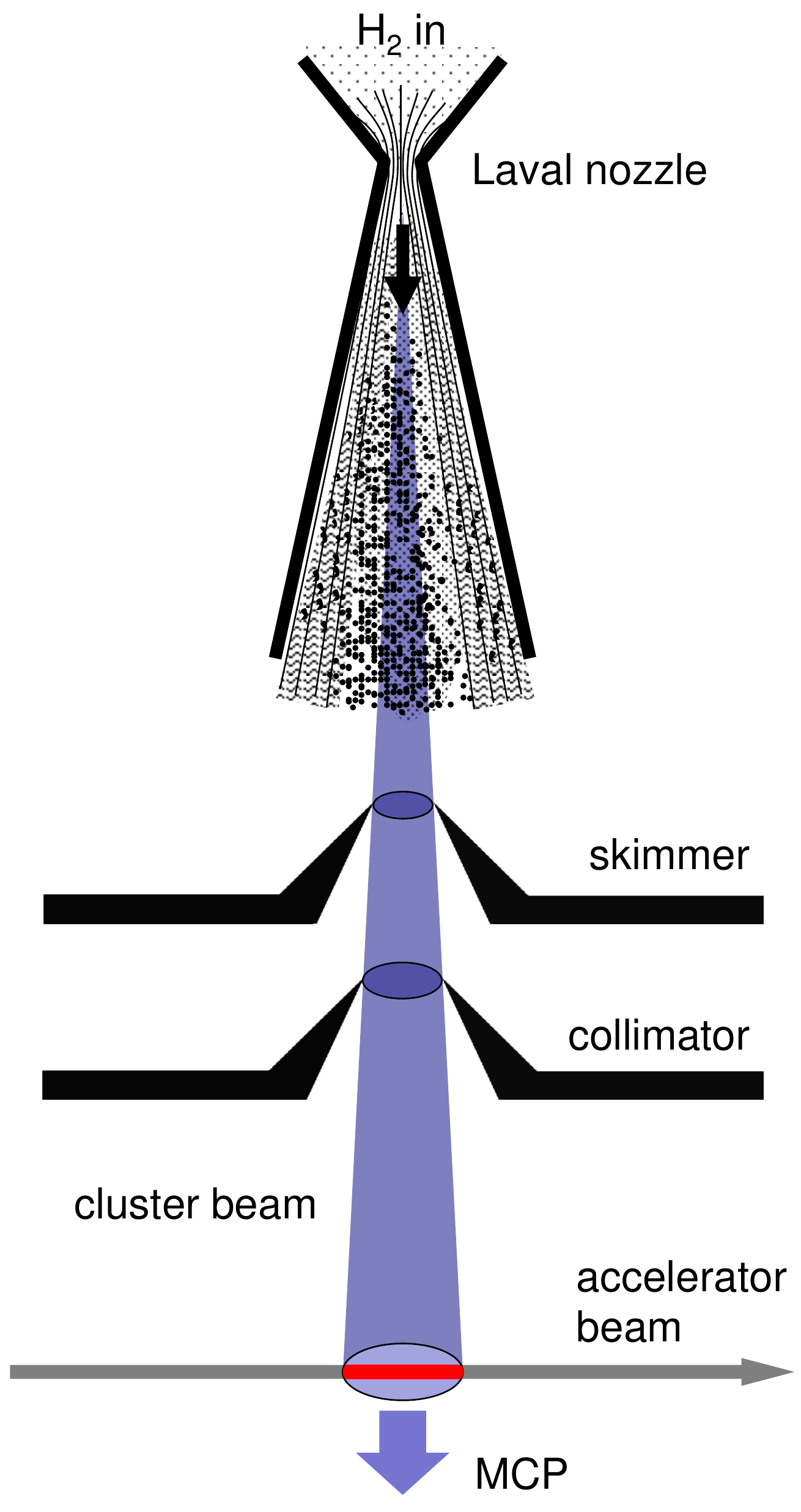}
\end{center}
\caption{Sketch of the partial ionization (red) of a cluster target beam by an accelerator beam.}
\label{fig:mcpcosy}
\end{figure}

The capability of the previously described MCP monitor device to detect  
spatially-resolved ionized clusters opens further possibilities for its use in 
combination with cluster beams as targets, e.g. in storage ring experiments. 
One valuable real time information which can be extracted quantitatively is the 
size of the vertex region, i.e. the interaction volume of the accelerated
ion beam with the cluster beam. A sketch of the operation 
principle is given in Fig.~\ref{fig:mcpcosy}. 

A cluster beam, produced in a Laval 
nozzle and prepared by a set of one skimmer and one collimator, enters the scattering
chamber of the particle accelerator and is hit by the ion beam. Due to energy loss processes, 
e.g. based on Coulomb effects, a certain fraction of the cluster beam of the overlap 
volume is ionized by the ion beam while the part of the cluster beam not hit by
the ions remains in an electrically neutral state. 
The partially ionized cluster beam enters behind the scattering chamber  
the beam dump stage where the previously 
described MCP monitor can be placed. 
Since only clusters from the interaction volume
are ionized, solely clusters from this region are expected to give a contribution 
to the resulting image.

\begin{figure}
\includegraphics[width=0.88\columnwidth]{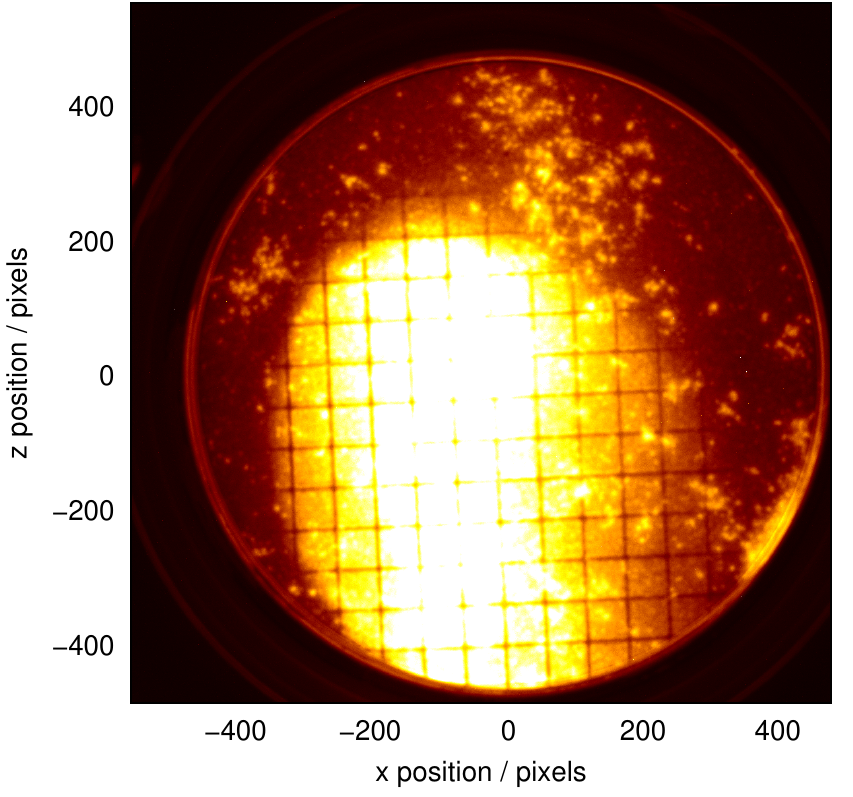}
\includegraphics[width=0.88\columnwidth]{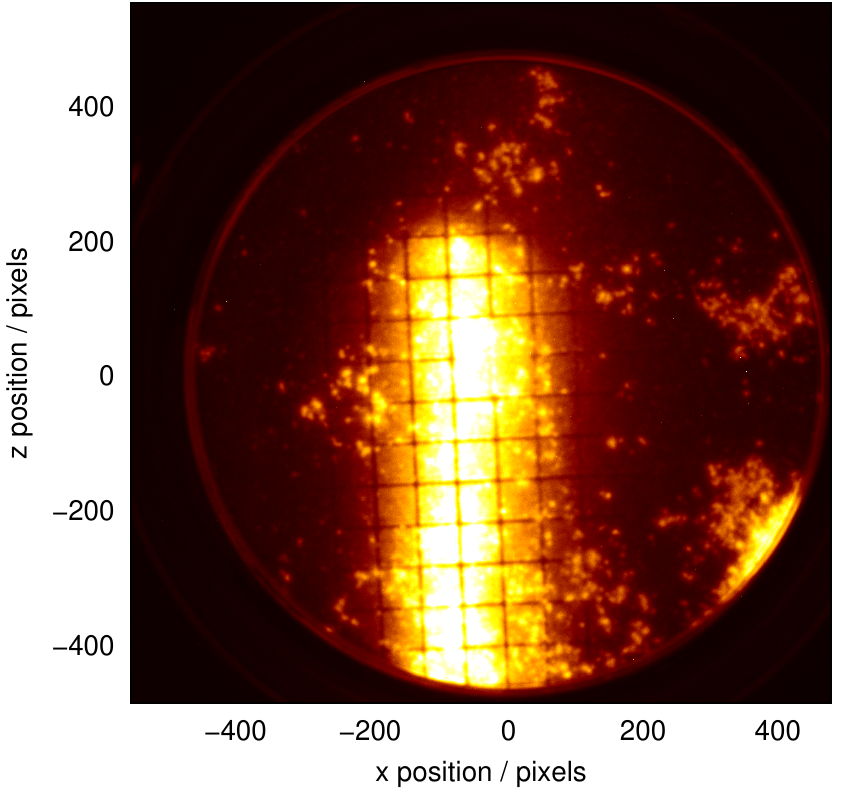}
\includegraphics[width=0.88\columnwidth]{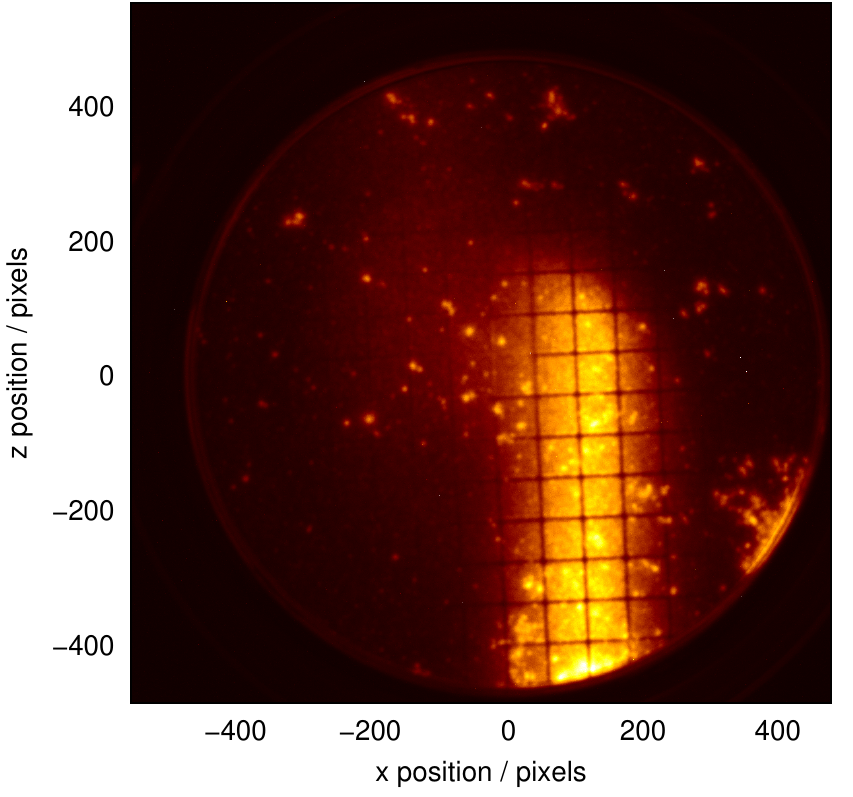}
\caption{Visualization of the beam-target vertex region during the acceleration of the COSY
beam (top) and at a flat top momentum of $2.1\,\mathrm{GeV/c}$ (center). The third figure
shows the vertex region also at $2.1\,\mathrm{GeV/c}$ but with a switched on dipole steering magnet 
in front of the scattering chamber. }
\label{fig:COSY_acc_steer}
\end{figure}

\begin{figure}[h]
\includegraphics[width=\columnwidth]{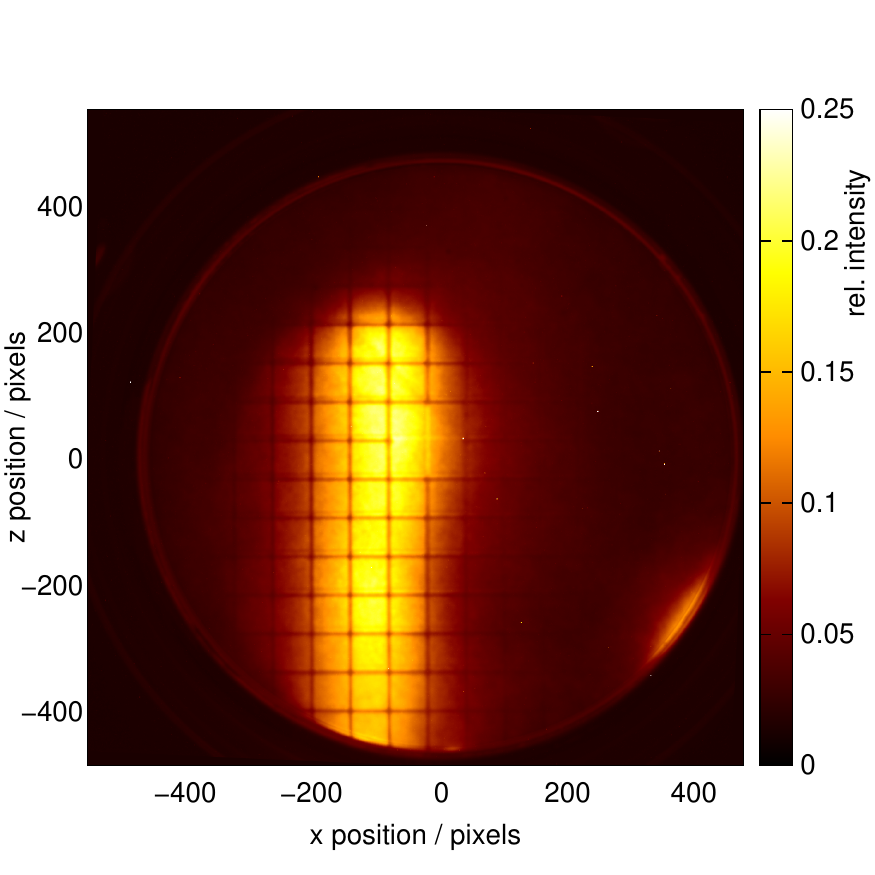}
\includegraphics[width=\columnwidth]{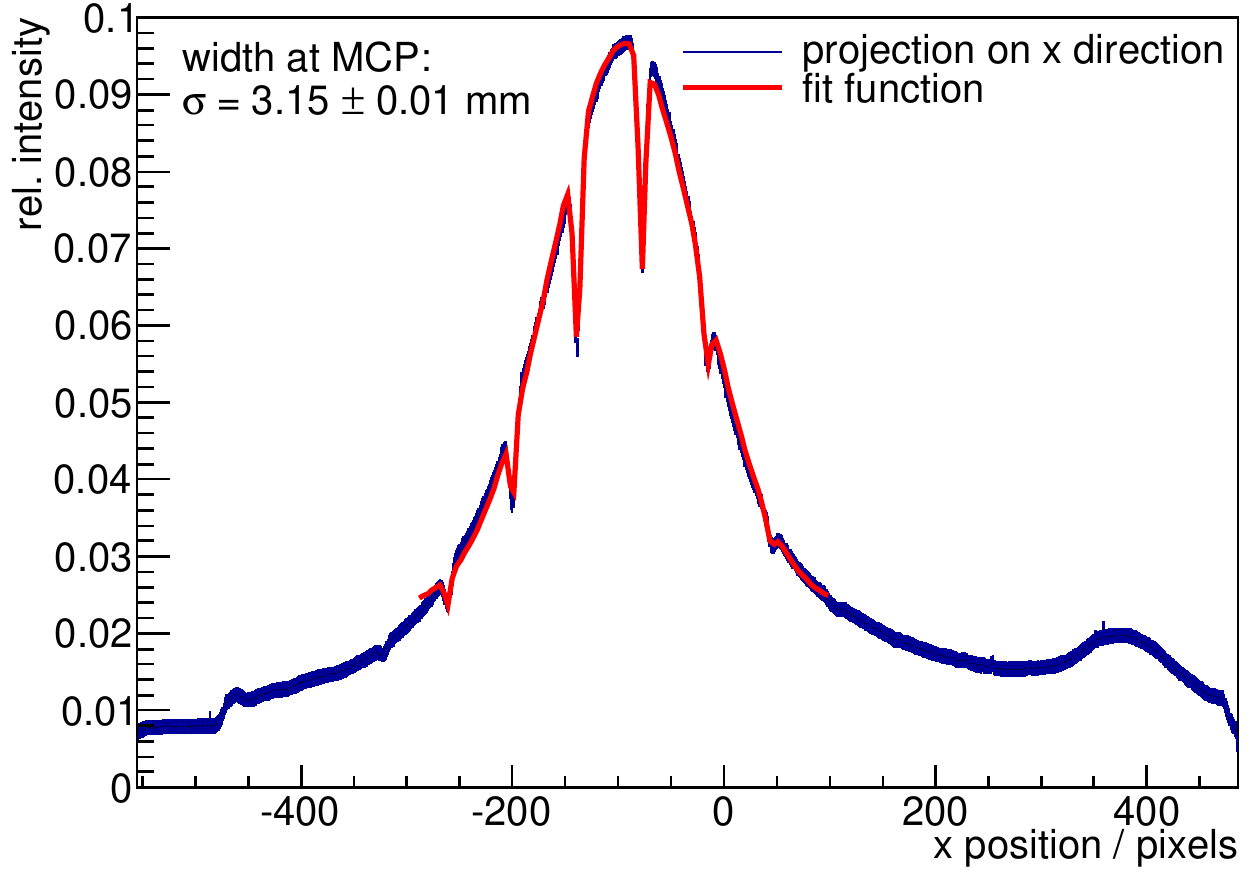}
\caption{Image and projection of the vertex region obtained by averaging 145 individual images
with an exposure time of $5\,\mathrm{s}$ each. }
\label{fig:COSY_proj}
\end{figure}

A feasibility study of this method has been performed \cite{GSIb} during a few hours of an
out of turn beam time 
at the ANKE experimental facility \cite{Bar01}
which is installed at an internal target position of the cooler synchrotron COSY \cite{Mai97} at the
Forschungszentrum J\"ulich. At ANKE a cluster jet target is installed which
provides hydrogen cluster target thicknesses of up to $\rho_{\mathrm{target}}=10^{15}\,\mathrm{atoms/cm}^2$
in a distance of $70\,\mathrm{cm}$ behind the nozzle 
in combination with a target diameter of approximately $10\,\mathrm{mm}$. For the studies presented 
here 
the final turbomolecular pump of the beam dump stage was replaced by the described MCP monitoring
device and a thin cluster jet beam was produced with $\rho_{\mathrm{target}}\le 10^{14}\,\mathrm{atoms/cm}^2$.
The COSY accelerator was operated in a mode where low energetic protons, injected with a momentum of 
$p_{\mathrm{inj}}=300\,\mathrm{MeV/c}$, where stored and accelerated up to a beam momentum
of $p_{\mathrm{acc}}=2100\,\mathrm{MeV/c}$. After a cycle length of $30\,\mathrm{s}$ the proton beam was 
dumped and new cycles with the same time structure and with newly injected protons followed. 
In Fig.~\ref{fig:COSY_acc_steer} (top) 
an image recorded with the MCP device is shown which was obtained during the acceleration of the
COSY beam. The exposure time of the CCD camera was set to $5\,\mathrm{s}$. A bright area is 
visible which corresponds to clusters ionized by the accelerator beam. The vertical axis (''z-position'') 
corresponds to the ion beam direction while the horizontal axis (''x-position'') corresponds to 
the one transverse to the beam direction. Note that the COSY beam has a comparably broad diameter during the acceleration
due to the low proton momentum shortly after the injection. In the second figure (Fig.~\ref{fig:COSY_acc_steer}, 
center) an image is shown which was recorded when the final COSY momentum was reached, i.e. $p_{\mathrm{acc}}=2100\,\mathrm{MeV/c}$. 
Due to the adiabatic cooling of the accelerator beam the transverse diameter of the ion beam shrunk which 
can be seen directly. In a next step one steering magnet in front of the ANKE scattering chamber was switched on
in order to allow for a shift of the COSY beam transverse to its flight direction. The resulting image 
is shown in Fig.~\ref{fig:COSY_acc_steer} (bottom) where a clear displacement is visible. The lower 
intensity is only an artifact due to a lower proton beam intensity during this measurement. 
Therefore, this series of measurements proofs that a direct observation of the interaction vertex is 
possible by such a MCP device. 
For a more quantitative investigation a series of 145 images recorded at $p_{\mathrm{acc}}=
2100\,\mathrm{MeV/c}$
has been analyzed and combined in one averaged image. The resulting image and a projection on the 
transverse axis is shown in Fig.~\ref{fig:COSY_proj}. Note that for the measurements presented 
here a grid was used where four joining grid bars in the center of this electrode where removed
(see Fig.~\ref{fig:COSY_proj}). However, this fact is of no relevance for the further discussion. 

Due to time limitations of the measurement the MCP device could not be 
aligned perfectly relative to the cluster beam axis, which resulted in the fact that the signal 
is not in the center of the MCP screen. 
Nevertheless, solid quantitative information about the 
vertex region can be gained with good accuracy.
The obtained data (blue line) can be described well by the assumption of a Gaussian intensity
distribution with a width of $\sigma_{\mathrm{COSY,MCP}} =(3.15\pm 0.01)\,\mathrm{mm}$ (RMS). Here again the 
well understood small intensity drops are caused by the grid structure, which can be treated
as described above. Since the cluster beam has a constant angular divergence given by the 
size of the collimator of the cluster source and its distance from the nozzle, the width of the 
COSY beam can be reconstructed by simple geometrical considerations. With the knowledge 
of the distance between the nozzle and the vertex point in the scattering chamber of 
$l_{\mathrm{vertex}} =700\,\mathrm{mm}$ and the distance between the nozzle and the MCP
of $l_{\mathrm{MCP}} =1829\,\mathrm{mm}$, the width of the COSY beam in the scattering chamber 
$\sigma_{\mathrm{COSY,vertex}}$ can be written as:
\begin{equation}
\sigma_{\mathrm{COSY,vertex}} = \sigma_{\mathrm{COSY,MCP}} \cdot \frac{l_{\mathrm{vertex}}}{l_{\mathrm{MCP}}}. 
\end{equation}
Thus the width of the COSY beam at the vertex point could be estimated to be approximately
$\sigma_{\mathrm{COSY,vertex}} =1.2\,\mathrm{mm}$ (RMS) which is in good agreement with the
expected accelerator beam size \cite{prasuhn}. Due to the not ideal experimental conditions 
no uncertainties are given here. However, based on the achieved results a precision for the 
determination of the vertex size is conservatively expected to be in the order of $150\, \mu\mathrm{m}$ (RMS).

\section{Extended applications}

In addition to the use of the MCP device presented here as a monitor for the time resolved, 
two-dimensional investigation of both the cluster beam thickness and the
vertex size distributions, further applications are possible. 
If an adjustable positive retardation potential $U_\mathrm{ret}$ is applied to the entrance 
grid instead of the grounding, this device works as an energy filter for the impinging 
ionized clusters with positive charge. 
Depending
on the stagnation parameters of the fluid before entering the Laval nozzle, the
produced hydrogen clusters typically have a mean speed of 
$\bar{v}_\mathrm{cluster} = 200-1000\,\mathrm{m/s}$. The width of the velocity distribution
$\Delta{v}_\mathrm{cluster}$
also depends on the stagnation parameters and was observed to be in the range of 
$\Delta{v}_\mathrm{cluster}/\bar{v}_\mathrm{cluster}$ = 0.5\% to 8\%.
For details see Ref.~\cite{Tas11,Koehler2010,TaeschnerDr}.
By investigating the intensity variation recorded by the CCD camera as a function of the 
retardation potential $U_\mathrm{ret}$, information about the masses 
of the ionized clusters $m_\mathrm{cluster}$ can be gained. In detail clusters 
of charge $q$ with a mass below
\begin{equation}
m_\mathrm{cluster} = \frac{2q U_\mathrm{ret}}{{v_\mathrm{cluster}}^2}
\end{equation}
can be reflected. This opens the possibility to investigate the mass distribution of 
cluster target beams, e.g. as function of the stagnation parameters or of the nozzle geometry,
in parallel to the operation as target in experiments. 
Assuming, for example, a retardation voltage of $U_\mathrm{ret}=1000\,\mathrm{V}$ and
a cluster velocity of $v_\mathrm{cluster} = 600\,\mathrm{m/s}$, clusters 
masses of up to $m_\mathrm{cluster} \approx 500.000\,\mathrm{amu}$ can be investigated.
Moreover, by comparing the obtained distributions with and without an accelerator beam
passing the cluster beam, information about a possible fragmentation of clusters 
caused by the ion beam can be 
gained. First pilot studies on cluster mass measurements with this device and using the described 
cluster target have been performed successfully and are followed by extensive systematic 
investigations \cite{Esperanza2013}.

\section{Summary}
In summary we have presented an advanced technique which enables, for the first time, 
the two-dimensional real time visualization of cluster
target beams as well as of vertex regions at internal beam experiments
at particle accelerators.
The system is based on a microchannel plate array in combination with a phosphor 
screen which is read out by a CCD camera. Full, time resolved, two-dimensional information about the 
cluster beam thickness distribution are accessible if a dedicated electron gun is switched on.
A spatial resolution in the order of 
$\sigma\,\approx\,100\,\mu\mathrm{m}$ has been reached which allows for quantitative investigations
on the size, shape and thickness distribution of the cluster beams itself. 
In addition it has been demonstrated that this device allows for a detailed 
two-dimensional investigation of the vertex region at storage ring experiments using internal 
cluster jet targets. Here the ionization of the clusters proceeds solely by the passing 
ion beam if the previously used electron beam is switched off. 
With this device a resolution of approximately $\sigma\,\approx\,150\,\mu\mathrm{m}$
was achieved. The method presented here can principally also be used for further applications, 
e.g. for laser-cluster interactions, where electrically charged clusters are produced, or
for the cluster mass investigation.

\section*{Acknowledgements}

The authors would like to thank the COSY crew for the outstanding support  
and for the possibility to perform the presented test measurements.
We acknowledge the excellent work done by our mechanical and electronic workshop. 
The research project was supported by BMBF (06MS253I, 06MS9149I/05P09MMFP8, 
06MS7190I/05P12PMFP5), GSI
F\&E program (MSKHOU1012), EU/FP6 HADRONPHYSICS (506078), EU/FP7
HADRONPHYSICS2 (227431) and EU/FP7 HADRONPHYSICS3 (283286).






\section*{References}



\end{document}